\documentclass[linenumber]{aastex631}

\usepackage{multirow}
\usepackage{amsmath}

\newcommand{\lyalpha}{Ly\,$\alpha$}

\newcommand{\NH}{neutral hydrogen}

\submitjournal{ApJL}

\shorttitle{Imaging magnetically driven astrospheres}
\shortauthors{Wu et al.}

\begin{document}
\title{Imaging magnetically driven astrospheres: a forward modelling approach}

\correspondingauthor{Ziqi Wu, Jiansen He}
\email{ziqi.wu@student.kuleuven.be, jshept@pku.edu.cn}

\author[0000-0002-1349-8720]{Ziqi Wu}
\affiliation{School of Earth and Space Sciences, Peking University \\
No.5 Yiheyuan Road, Haidian District,
Beijing, 100871, China}
\affiliation{Centre of mathematical Plasma Astrophysics, Department of Mathematics, KU Leuven \\ 
Celestijnenlaan 200B, 3001 Leuven, Belgium}

\author[0000-0001-9628-4113]{Tom Van Doorsselaere}
\affiliation{Centre of mathematical Plasma Astrophysics, Department of Mathematics, KU Leuven \\ 
Celestijnenlaan 200B, 3001 Leuven, Belgium}

\author[0000-0001-8179-417X]{Jiansen He}
\affiliation{School of Earth and Space Sciences, Peking University \\
No.5 Yiheyuan Road, Haidian District,
Beijing, 100871, China}

\author[0000-0001-6656-4130]{Hugues Sana}
\affiliation{Institute of Astronomy, KU Leuven\\
Celestijnenlaan 200D, 3001 Leuven, Belgium}

\author[0000-0003-4670-9616]{Nicholas Jannsen}
\affiliation{Institute of Astronomy, KU Leuven\\
Celestijnenlaan 200D, 3001 Leuven, Belgium}
\affiliation{Isaac Newton Group of Telescopes (Roche de Los Muchachos)\\ 
Apartado 321, E-38700 Santa Cruz de La Palma, Canaries, Spain}

\author[0000-0001-7170-0408]{Tianhang Chen}
\affiliation{School of Earth and Space Sciences, Peking University \\
No.5 Yiheyuan Road, Haidian District,
Beijing, 100871, China}

\author[0009-0006-0032-0700]{Weining Wang}
\affiliation{School of Earth and Space Sciences, Peking University \\
No.5 Yiheyuan Road, Haidian District,
Beijing, 100871, China}

\author[0000-0001-5657-7587]{Zheng Sun}
\affiliation{School of Earth and Space Sciences, Peking University \\
No.5 Yiheyuan Road, Haidian District,
Beijing, 100871, China}
\affiliation{Leibniz Institute for Astrophysics Potsdam\\
An der Sternwarte 16, 14482 Potsdam, Germany}

\begin{abstract}
An astrosphere is a vast, tailed bubble-like volume around a star, formed through the interaction between the stellar magnetic field, the stellar wind, and the interstellar medium (ISM). Detecting and characterizing astrospheres are essential for constraining stellar wind properties, understanding stellar evolution, and assessing the habitability of surrounding exoplanetary systems. Charge exchanges between ionized stellar wind particles and cold ISM hydrogen atoms populate the astrosphere with neutral hydrogen, which can leave observable signatures in the Lyman-$\alpha$ (\lyalpha{}) line absorption profile. Previous studies have inferred stellar mass-loss rates by measuring \lyalpha{} absorption in stellar spectra caused by astrospheric neutral hydrogen. However, our knowledge of the global morphology of astrospheres remains limited and largely dependent on sometimes contradictory simulations.
Here we investigate the feasibility of detecting \lyalpha{} emission generated by resonant scattering from \NH{} surrounding the star, enabling the construction of a two-dimensional map of the astrosphere. With a three-dimensional magnetohydrodynamic astrosphere model, we perform forward modelling of the \lyalpha{} emission and assess the observation feasibility according to the observational limits of the {\it Hubble Space Telescope} (HST). We further discuss the influence of varied line-of-sight orientations and averaged ISM velocity along the line-of-sight. The spatially resolved circumstellar \lyalpha{} emission could provide important constraints on the astrospheric configuration and stellar wind properties, such as the bow shock standing distance, the stellar wind symmetry, and the shape of the astro-tail. Our results highlight \lyalpha{ astrosphere detections as a promising science case for {\it HST} and future missions such as the \textit{Habitable Worlds Observatory}.}
\end{abstract}

\section{Introduction}
Stellar wind is plasma outflow from a star into its surrounding space \citep{Lamers&Cassinelli1999}. This wind carves out an astrosphere where exoplanetary systems reside in the interstellar medium (ISM)   \citep{Dong2017, RodriguezMozos&Moya2019, Canet2024}. They also transport mass, energy, and angular momentum throughout the astrosphere, causing mass loss and spin-down of the host star \citep{Brott2011, Gallet&Bouvier2013, Gallet&Bouvier2015}. While the solar wind has been studied intensively for decades since its prediction by \citet{Parker1958}, stellar winds in general remain less well constrained. Studying stellar winds is essential for predicting the habitability of exoplanetary systems and understanding the early history of the Sun \citep{Vidotto2021}.

Estimating the mass flux of the stellar wind is a primary goal in stellar wind studies. Stellar mass-loss rates vary widely with stellar type and evolutionary stage, ranging from as low as $2\times10^{-14} ~M_{\odot}\mathrm{yr}^{-1}$ for the Sun based on solar wind observations \citep{Cohen2011} on the main sequence to as high as $10^{-6}$-$10^{-5}~M_{\odot}\mathrm{yr}^{-1}$ for asymptotic giant branch stars \citep{Hofner&Olofsson2018}. For stars with strong winds, such as massive hot stars, giant stars, and T Tauri stars, mass-loss rates have been estimated using observations of radio emission, optical or near-infrared recombination lines (e.g., H I, He II 4686~$\AA$), and ultraviolet spectroscopy \citep{Bieging&Churchwell1982, Scuderi1998, Drake1986, Cohen1986, Bouret2013, Hawcroft2024, Brunella2024}.

For low-mass main-sequence stars, however, the ionized stellar wind is extremely difficult to detect directly. Such observational efforts to detect ionized stellar winds include X-ray \citep{Wargelin&Drake2002, Kislyakova2024, Lisse2026} and radio observations \citep{Gaidos2000, Fichtinger2017, Bloot2025}. An indirect detection method relies on charge-exchange interactions between stellar wind protons and neutral atoms or molecules in the ISM \citep{Cravens1997, Cravens2000}. During a charge exchange collision, a stellar wind proton captures an electron from a neutral particle, producing a pick-up ion and a fast neutral atom, commonly referred to as an energetic neutral atom. Energetic neutral atoms largely retain the original velocity of the parent stellar wind protons and therefore propagate faster than the typical ISM \NH.

\begin{figure}[htbp]
    \centering
    \includegraphics[width=0.8\linewidth]{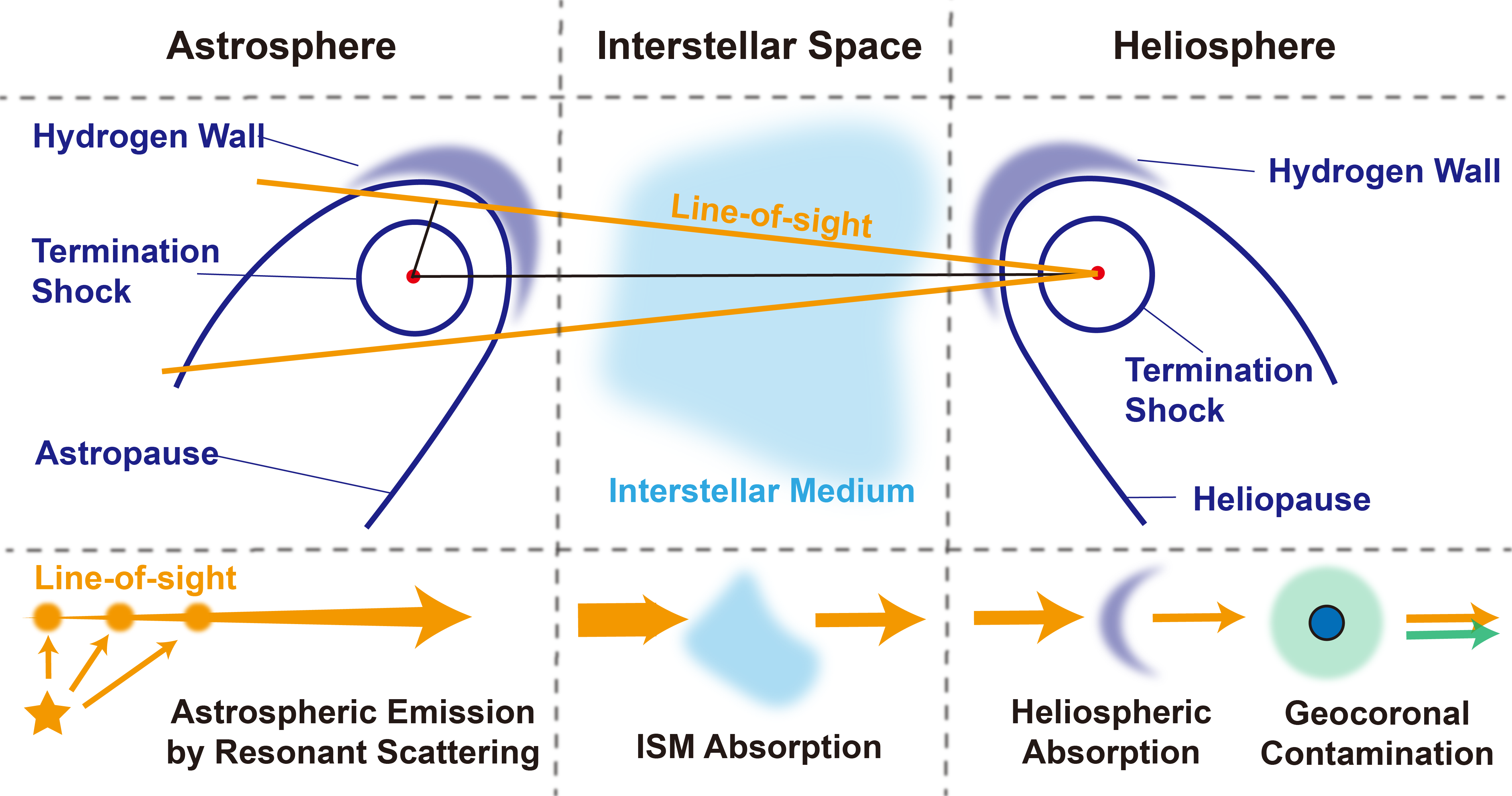}
    \caption{Illustration of the generation and propagation of the circumstellar \lyalpha{} emission.}
    \label{fig: cartoon}
\end{figure}

The \NH{} generated through charge exchange between interstellar \NH{} and the stellar wind between the heliopause and the bow shock/wave accumulates to form a dense hydrogen wall (blue shaded regions near the astropause/heliopause in Figure~\ref{fig: cartoon}). Using {\it Hubble Space Telescope} (HST) observations, \citet{Wood2004} identified excess absorption in the blue wing of stellar Lyman-$\alpha$ (\lyalpha{}) profiles and attributed it to \NH{} in the stellar hydrogen wall. Subsequent observations demonstrated that several nearby stars exhibit similar astrospheric absorption features \citep{Wood2005, Wood2005b, Wood2021}. By comparison with two-dimensional (2D) hydrodynamic (HD) models, the stellar mass-loss rates were estimated. The results revealed that stellar mass-loss rates generally increase with increasing stellar activity, with notable exceptions among GK dwarfs. For stars exhibiting high coronal activity but relatively weak winds, magnetic topology has been proposed as a key factor regulating wind outflow \citep{Wood2021}. For a sample of 21 cool main-sequence stars (from F-type to M-type dwarfs) compiled by \citet{Wood2021}, magnetized stellar winds have also been simulated using state-of-the-art three-dimensional (3D) magnetohydrodynamic (MHD) models driven by observed large-scale magnetic field distributions \citep{Chebly2023}. \citet{KorolkovandIzmodenov2026} further examined the absorption due to \NH{} generated by charge exchange between the supersonic stellar wind protons and ISM atoms with different stellar wind terminal velocities.

The neutral hydrogen may also provide an indirect means of probing stellar winds by producing observable signatures in the astrospheric \lyalpha{} emission profile. The \NH{} throughout the astrosphere contributes to \lyalpha{} emission by resonantly scattering photons from the host star. Taking the heliosphere as an example, \NH{} emission and absorption primarily generated from four regions: the local ISM, the compressed plasma region downstream of the bow shock, the inner heliosheath, and the supersonic solar wind \citep{Malama2006, Izmodenov2015, Opher2020}. The \NH{} in all of these regions can, in principle, absorb and re-emit \lyalpha{} photons through resonant scattering \citep{Ouchi2020}, potentially producing observable emission signatures (see Figure~\ref{fig: cartoon}). Owing to their higher bulk velocities and temperature relative to the ISM, neutral hydrogen generated by charge exchange can produce emission in the wings of the \lyalpha{} profile. Such emission is more likely to survive ISM absorption, heliospheric absorption, and geocoronal contamination during propagation (see Figure~\ref{fig: cartoon}). Recombination followed by cascade can also produce Lyman-alpha emission, as electrons recombine with protons into excited states and subsequently decay through the 2p level \citep{Hirata2009, Hayes2019}. However, because the emissivity scales with electron and proton densities and decreases with increasing temperature, this process is negligible in the hot, tenuous solar wind compared to resonant scattering. It may nevertheless contribute in denser, cooler environments such as exoplanetary atmospheres \citep{Linsky2024}.

Besides providing information on stellar mass-loss rates, circumstellar \lyalpha{} emission also offers a way to probe the morphology of astrospheres. In the heliosphere, the Interstellar Boundary Explorer mission has directly detected energetic neutral atoms produced through charge exchange at the heliospheric boundaries, providing important constraints on its global structure \citep{Schwadron2009, McComas2009, McComas2024}. Yet the overall shape of the heliosphere remains under debate, with different models suggesting single tail (comet-like), tailless (potato-like), or double tail (croissant-like) geometries \citep{Pogorelov2015, Pogorelov2025, Opher2015, Opher2020}. The large-scale structure of heliotail is governed by the interaction between the solar wind and the ISM, involving magnetic tension–driven solar wind jets, hot pick-up ions, and the surrounding interstellar flow. For other stars, astrospheres may in fact be easier to characterize morphologically, since we observe them from the outside. Understanding their large-scale structure can therefore help constrain the properties of stellar winds and magnetic fields, as well as diagnose the presence of large-scale waves and instabilities.

\citet{Wood2003} conducted a tentative observation with the {\it HST} of \lyalpha{} emission scattered by the hydrogen wall of a nearby star, 40~Eri~A (distance: 5.0 pc), but reported no significant detection. Nevertheless, modeling efforts based on a 2D asymmetric four-fluid HD model of the 40~Eri~A astrosphere, coupled with Monte Carlo radiative transfer simulations, indicated that such emission should exist. This apparent discrepancy was attributed to the possibility that 40~Eri~A resides outside the relatively cold local interstellar cloud and is instead embedded in a hot, highly ionized ISM, which would significantly reduce the \NH{} density and suppress \lyalpha{} scattering. Resolving this puzzle and extending \lyalpha{} emission studies to a broader sample of stars will require additional observations as well as more sophisticated astrospheric models.

In this study, we investigate the feasibility of mapping astrospheres in 2D by analyzing the circumstellar \lyalpha{} emission. As a first step, we employ a 3D MHD heliosphere model embedded in a near-Sun ISM environment and forward-model the resulting \lyalpha{} emission and observable intensity, explicitly accounting for ISM absorption and Doppler shifts. We further discuss the effects of heliospheric absorption and geocoronal emission, and assess the detectability of these signatures with HST. If successful, such 2D observations of astrospheres could provide essential constraints on key astrospheric properties, including the stellar mass-loss rates, bow-shock standoff distance, and astrosphere morphology such as astrotail shapes.

\section{Data and Method}
\subsection{Astrosphere model}
To forward model the scattered \lyalpha{} emission from the astrosphere, we construct a 3D MHD astrosphere model that self-consistently describes the magnetic field, density, velocity, and temperature of multiple particle species, including low-energy thermal ions, pick-up ions, and \NH{} particles. The model corresponds to the Outer Heliosphere component of the Space Weather Modeling Framework (SWMF), which employs the 3D MHD code Block-Adaptive Tree Solarwind Roe-type Upwind Scheme (BATS-R-US) with adaptive mesh refinement \citep{Toth2012}. The \NH{ distribution includes four populations: (1) hydrogen generated by charge exchange downstream of the bow shock, (2) hydrogen generated downstream of the termination shock, (3) hydrogen generated in the supersonic stellar wind, and (4) interstellar hydrogen \citep{Opher2015, Opher2020}.}

To simulate the astrosphere, we adopt the model setting used in \citet{Opher2020} for the heliosphere. The lower boundary of the model is set at 30~AU, where the stellar wind parameters are aligned with that of solar wind protons, fixed at number density $n_{\mathrm{SW}} = 8.74 \times 10^{-3}~\mathrm{cm^{-3}}$, temperature $T_{\mathrm{SW}} = 1.09 \times 10^{5}~\mathrm{K}$, and radial bulk velocity $v_{\mathrm{SW}} = 417~\mathrm{km~s^{-1}}$. The magnetic field is prescribed using the Parker spiral model, with the radial component fixed at $B_r = 7.17 \times 10^{-1}~\mathrm{nT}$ at $30~\mathrm{AU}$. A monopole magnetic field is employed to avoid artificial magnetic reconnection. 

The upper boundary of the simulation domain is set at $x = \pm 1500~\mathrm{AU}$, $y = \pm 2000~\mathrm{AU}$, and $z = \pm 2000~\mathrm{AU}$. At this boundary, the ionized component of the ISM is prescribed with uniform number density $n_{\mathrm{ISM}} = 9.8 \times 10^{-2}~\mathrm{cm^{-3}}$, temperature $T_{\mathrm{ISM}} = 7500~\mathrm{K}$, bulk velocity $v_{\mathrm{ISM}} = 25.4~\mathrm{km~s^{-1}}$. The ISM inflow direction is $\sim75^{\circ}7$ in ecliptic longitude, $\sim5^{\circ}1$ in ecliptic inflow latitude \citep{McComas2015}. The interstellar magnetic field has magnitude of $2.93\pm 0.08~\mathrm{\mu G}$, with a direction of $227^{\circ}28 \pm 0^{\circ}69$ in ecliptic longitude and $ 34^{\circ}62 \pm 0^{\circ}45$ in ecliptic latitude \citep{Zirnstein2016}. The \NH{} in the ISM has a density of $n_{\mathrm{H, ISM}}=0.154~\mathrm{cm^{-3}}$ and shares the same velocity and temperature as the ions.

\begin{figure}
    \centering
    \includegraphics[width=0.75\linewidth]{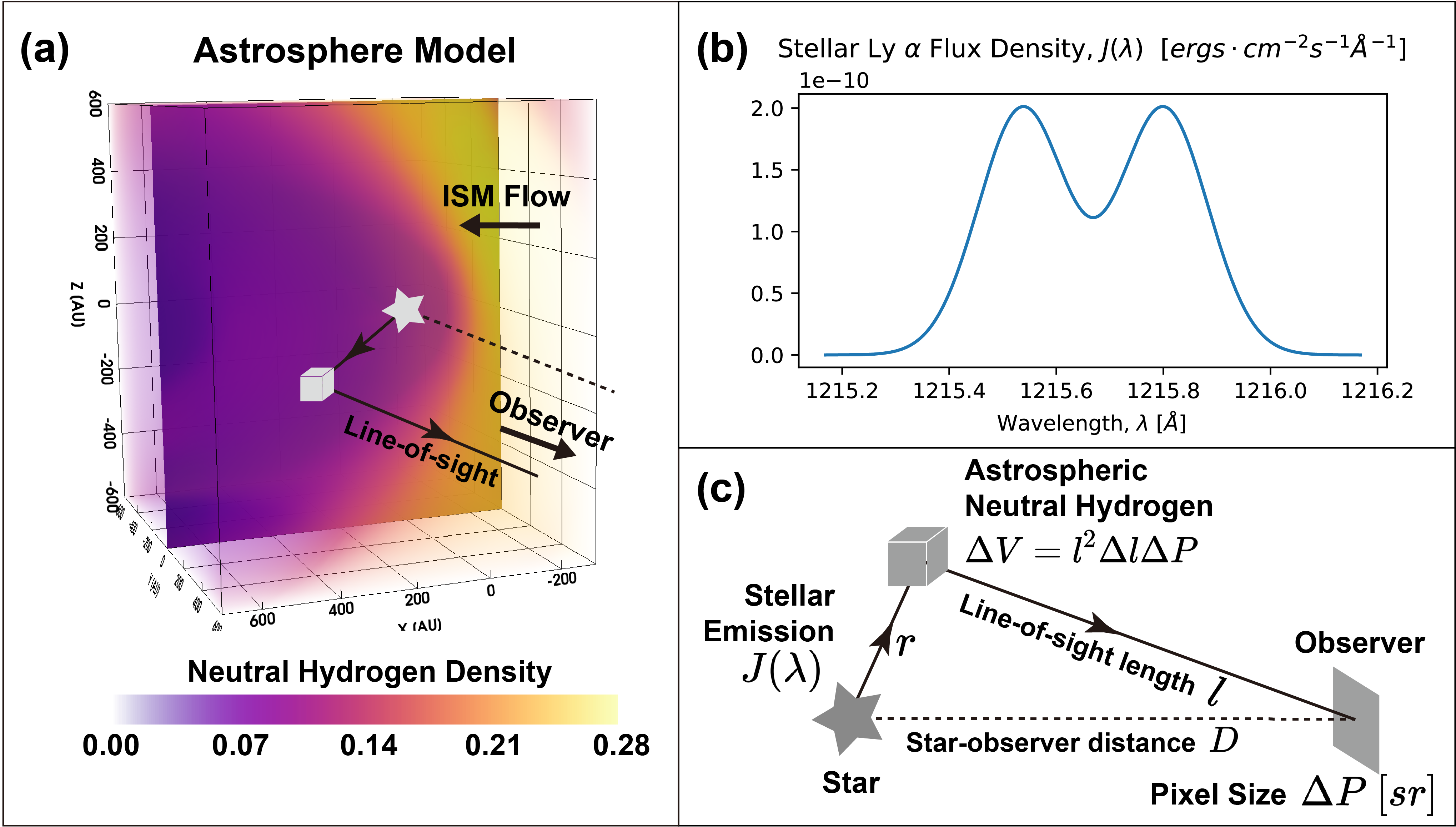}
    \caption{The stellar and astrosphere model and observing geometry. (a) The modeled distribution of \NH{} number density in the astrosphere. The directions of upwind and the observer are marked by arrows. (b) The modeled \lyalpha{} flux density profile of the central Star. (c) An illustration of the observing geometry.}
    \label{fig: OH_model}
\end{figure}

The coordinate system of the model is designed so that the $Z$ axis is parallel to the solar rotation axis, the $X$ axis is oriented in the direction of the interstellar flow (which points 5° upward in the $X$–$Z$ plane), and the $Y$ axis completes the right-handed coordinate system. We put this astrosphere model 1 pc away along the $Y$ axis from an observer located at $(0, -1 ~\mathrm{pc}, 0)$. Therefore, all the lines of sight from the observer are roughly parallel to the $Y$ axis. We focus on the region within $-300~\mathrm{AU} < x <700~\mathrm{AU}$, $-600~\mathrm{AU} < y  <600~\mathrm{AU}$, and $-600~\mathrm{AU} < z  <700~\mathrm{AU}$. Figure~\ref{fig: OH_model}(a) shows the modeled number density distribution of \NH{} and Figure~\ref{fig: OH_model}(c) illustrates the observing geometry. 

\begin{figure}
    \centering
    \includegraphics[width=0.8\linewidth]{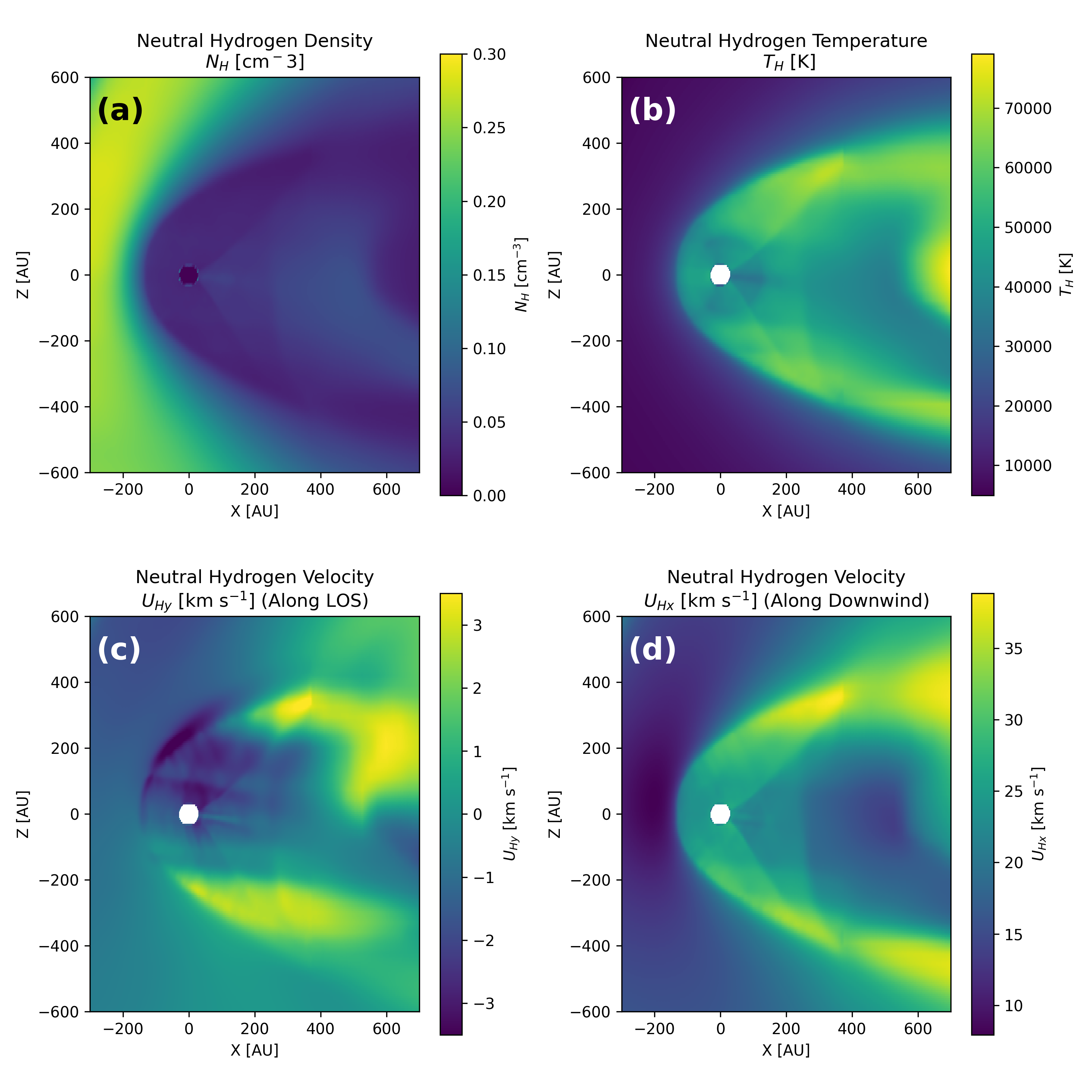}
    \caption{Slices of the astrosphere model. (a) Distribution of \NH{} number density. (b) Distribution of the \NH{} temperature. (c) Distribution of the \NH{} speed along the line-of-sight. (d) Distribution of the \NH{} speed along the downwind direction. The 3D astrosphere model is sliced at the $X$-$Z$ plane, where $X$ is the downwind direction, and $Z$ is aligned with the star's rotational axis. The interstellar flow comes from the left, and the white symbol represents the inner region below the lower boundary at 30~AU.}
    \label{fig: 2d_astrosphere}
\end{figure}

Figure~\ref{fig: 2d_astrosphere} shows the properties of \NH{} on a meridional slice of the astrosphere model. The \NH{} properties are averaged over all \NH{} populations, weighted by their number densities. As a result, these properties are dominated by the \NH{} generated by charge exchange between stellar wind and ISM behind the bow shock, which has the highest number density. The \NH{} reaches its highest density in the hydrogen wall near the bow shock, and is also enhanced in the astro-tail in the downwind direction (see Figure~\ref{fig: 2d_astrosphere}(a)). Inside the astrosphere, the \NH{} is significantly hotter, particularly near the astrospheric boundary and within the astro-tail (see Figure~\ref{fig: 2d_astrosphere}(b)). The velocity component perpendicular to the meridional plane (along the line-of-sight) remains relatively small, ranging from -3 to 3~km~s$^{-1}$ (Figure~\ref{fig: 2d_astrosphere}(c)). Along the downwind direction, \NH{} flows faster inside the astrosphere than in the ISM. These fast and hot \NH{} populations inside the astrosphere may contribute to observable \lyalpha{} absorption features.

We model the host star's \lyalpha{ profile, $J(\lambda)$ (Figure~\ref{fig: OH_model}(b)) with a double-Gaussian fit, aiming to mimic the double-peaked \lyalpha{} profile observed in both solar \citep{Gunar2020} and stellar  \citep[such as $\alpha$~Cen][]{Linsky1996} spectral flux density observed at Earth. The ISM absorption is not included in this profile}:

\begin{align}
    J(\nu) &= J_0\exp{\left[-\frac{(x-\Delta x)^2}{2}-\frac{(x+\Delta x)^2}{2}\right]}\\
    x &=\frac{\nu-\nu_c}{v_{\mathrm{th}}\nu_c/c},
\end{align}

where $J_0=2.0\times10^{-10} \mathrm{ergs}~\mathrm{cm}^{-2}\mathrm{s}^{-1}\mathrm{\AA}^{-1}$ is the maximum solar \lyalpha{} emission intensity observed at 1~pc away, $\nu$ is the frequency, $\nu_c=2.46605\times10^{15}~\mathrm{Hz}$ is the central frequency of \lyalpha{}, $v_{\mathrm{th}}$ is the thermal speed of protons corresponding to a temperature of $T = 2.5\times10^4~\mathrm{K}$, $c$ is the speed of light and $\Delta x=1.6$ is the dimensionless parameter indicating shifts of emission peaks.


\subsection{Radiative Transfer Calculation}
We perform \lyalpha{} radiative-transfer calculations \citep{Verhamme2006, Ouchi2020}, while accounting for the Doppler shifts introduced by the stellar wind velocity field. The geometry of the calculation is illustrated in Figure~\ref{fig: OH_model}(c). The star-observer distance is $D$. For each image pixel on the observer's imaging plane, subtending a solid angle $\Delta P$, we trace a single line-of-sight through the astrosphere. For a cell along the line-of-sight at length $l$ from the observer, the volume is $\Delta V=l^2\Delta l\Delta P$. The \NH{} in this cell absorbs \lyalpha{} photons from the star and re-emits them isotropically towards the observer. The pixel size is set as $\Delta P=(0.25 ~\mathrm{arcsec})^2\approx1.47\times10^{-12}~\mathrm{sr}$, corresponding to that of the Space Telescope Imaging Spectrograph (STIS) instrument onboard {\it HST} \citep{STIShandbook2024}.

For a given cell $\Delta V$ containing \NH{} with number density $n_H^{(i)}$ and temperature $T_H^{(i)}$, where the superscript $(i)$ denotes one of the four \NH{} populations, the \lyalpha{} absorption cross section is obtained by convolving the single-atom cross section with the Maxwellian velocity distribution of the gas. The resulting cross section is

\begin{equation}
    \sigma_x^{(i)}(\nu, T_H^{(i)})=\frac{3 \lambda_0^2 a_v}{2 \sqrt{\pi}} H\left(a_v, x\right) \simeq 5.9 \times 10^{-14}\left(\frac{T_H^{(i)}}{10^4 \mathrm{~K}}\right)^{-1 / 2} H\left(a_v, x\right) \mathrm{cm}^2,
    \label{crosssection}
\end{equation}
where $\lambda_0$ is the line center wavelength, $a_v = 4.7 \times 10^{-4}\left(T_{H}^{(i)}/10^4 \mathrm{~K}\right)^{-1 / 2}$ is the Voigt parameter for \lyalpha{}, $x$ is a dimensionless frequency defined as $x \equiv\left(\nu-\nu_0\right) / \Delta \nu_D$, where $\nu_0$ is the line center frequency and $\Delta\nu_D$ is the Doppler width. The Voigt function $H(a_v, x)$ is defined as:

\begin{equation}
    H\left(a_v, x\right)=\frac{a_v}{\pi} \int_{-\infty}^{\infty} \frac{e^{-y^2} d y}{(y-x)^2+a_v^2} \approx \begin{cases}e^{-x^2} & \mathrm { central\ resonant\ core } \\
\frac{a_v}{\sqrt{\pi} x^2} & \mathrm {damping\  wing.}
\end{cases}
\end{equation}

The transition between the core and the wing lies at $e^{-x^2}=a_v/(\sqrt{\pi}x^2)$.

The \lyalpha{} energy absorbed by the  $i$-th \NH{} population in this cell is 

\begin{equation}
E_{\mathrm{absorb}}^{(i)}=J(\nu) \cdot \frac{D^2}{r^2} \sigma(\nu) n_H^{(i)} \Delta V=J(\nu) \cdot \frac{D^2}{r^2} \sigma(\nu) n_H^{(i)} \Delta l \Delta P l^2,
\end{equation}
where $J(\nu)$ is the modelled stellar \lyalpha{} emission. Because $J(\nu)$ is fitted to the \lyalpha{} spectral flux density of the host star as observed at Earth, we scale it by $\frac{D^2}{r^2}$ to obtain the flux incident on the volume element at distance $r$ from the star.

Assuming isotropic re-emission, the fraction of the absorbed energy scattered towards the observer is
\begin{equation}
E_{\mathrm{scatter}}^{(i)}=J(\nu) \cdot \frac{D^2}{r^2} \sigma(\nu) n_H^{(i)} \Delta l \Delta P l^2 /\left(4 \pi l^2\right)=J(\nu) \cdot \frac{D^2}{4\pi r^2} \sigma(\nu) n_H^{(i)} \Delta l \Delta P.
\end{equation}

To properly account for the Doppler shift experienced between the emitting cell and the observer, the scattering function is further convolved with a Gaussian velocity distribution along the line-of-sight. For each \NH{} population, the distribution is parametrized by their bulk velocity $u_{los}^{(i)}$ and thermal velocity $v_{th}^{(i)}$ separately. The turbulent broadening is generally much smaller than the thermal velocity and is therefore not included.

The total \lyalpha{} surface brightness along a given line-of-sight is then obtained by integrating $E_{\mathrm{scatter}}^{(i)}$ over the entire light-of-sight length $l$ through the astrosphere and summing up contributions from all four \NH{} populations.

Finally, we calculate the ISM absorption from the star to the observer. We use the transmission rate to account for the ISM absorption by neutral hydrogen and deuterium \citep{Wood2005, Youngblood2016}: 

\begin{equation}
    \mathcal{T}^{(\mathrm{H,~D})}(\nu)=\exp{\left(-N_{\mathrm{ISM}}^{(\mathrm{H,~D})}\cdot \sigma\left(\nu(v_{\mathrm{ISM}}^{(\mathrm{H,~D})},b_{\mathrm{ISM}}^{(\mathrm{H,~D})}), T_{\mathrm{ISM}}^{(\mathrm{H,~D})}\right)\right)},
\end{equation}
where $N_{\mathrm{ISM}^{(\mathrm{H,~D})}}$ is the column density, $T_{\mathrm{ISM}}^{(\mathrm{H,~D})}$ is the ISM temperature, $\sigma$ is the cross section calculated with Equation \ref{crosssection} where line-of-sight speed $v_{\mathrm{ISM}}^{(\mathrm{H,~D})}$ and Doppler width $b_{\mathrm{ISM}}^{(\mathrm{H,~D})}$ are also included. For hydrogen, we use parameters as in \citet{Redfield&Linsky2000}, where $N_{\mathrm{ISM}}^{(\mathrm{H})}=n_{\mathrm{ISM}}^{(\mathrm{H})}\cdot D=0.1~\mathrm{cm}^{-3} \cdot 1~\mathrm{pc} \approx 3\times 10^{17}~\mathrm{cm}^{-2}$, $T_{\mathrm{ISM}}^{(\mathrm{H})}=7000~\mathrm{K}$. For deuterium, the density ratio $\mathrm{D/H}$ is $1.5\times10^{-5}$, constant within the Local Bubble \citep{Wood2005b, Linsky2006}, the centroid velocity $v_{\mathrm{ISM}}^{(\mathrm{D})}$ and temperature $T_{\mathrm{ISM}}^{(\mathrm{D})}$ are the same as hydrogen, the Doppler width follows $b_{\mathrm{ISM}}^{(\mathrm{D})}=b_{\mathrm{ISM}}^{(\mathrm{H})}/\sqrt{2}$. The central wavelength of deuterium absorption is $-0.33\AA$ from that of hydrogen absorption.

\section{Results}
\begin{figure}[htbp]
    \centering
    \includegraphics[width=0.6\linewidth]{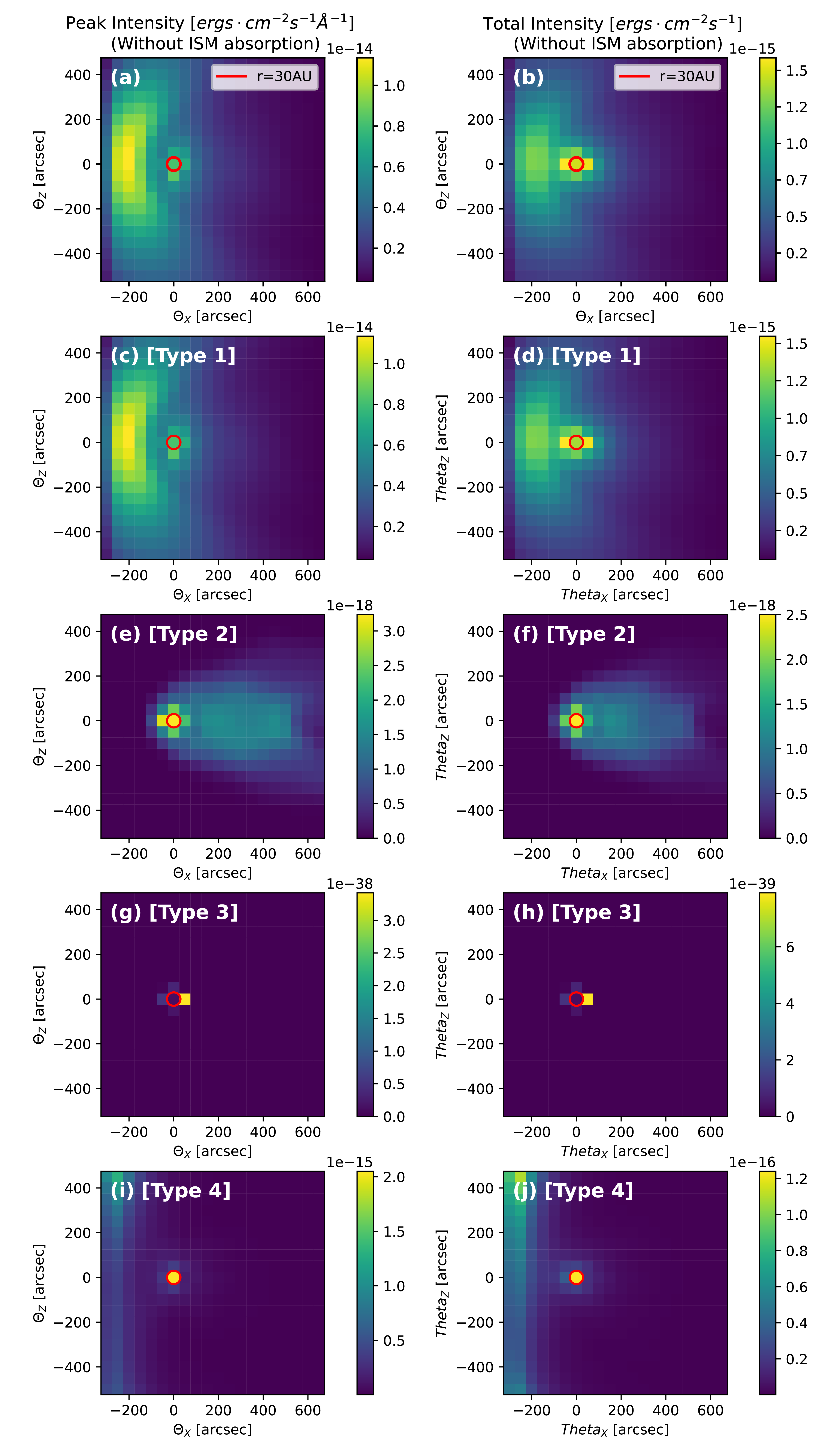}
    \caption{Modeled \lyalpha{ emission from the astrosphere. The left column shows the distribution of peak intensity, the right column shows the emission intensity integrated over all wavelengths. The red circle represents the bottom boundary of the heliosphere model located at $30~\mathrm{AU}$. The first row (a,b) shows the combined emission from all four populations. Other rows show the separate contribution of neutral hydrogen (c,d) generated downstream of the bow shock (Type 1); (e,f) generated downstream of the termination shock (Type 2); (g,h) generated in the supersonic stellar wind (Type 3); (i,j) originated from the ISM (Type 4).}}
    \label{fig: emission_without_ISM_absorption}
\end{figure}

With the astrosphere model and radiative transfer calculation described above, we model the astrospheric \lyalpha{} emission observed from 1 pc away. The boresight of the modeled image is centered on the star, and $\Theta_X$ and $\Theta_Z$ denote angular offsets from the boresight of the modelled grid, with each 1 arcsec corresponding to 1 AU in the plane of the sky, given the distance $D=1$~pc to the fiducial target. For each line-of-sight at ($\Theta_X$, $\Theta_Z$), we compute the \lyalpha{} emission spectrum. Note that the resolution of the modelled grid does not correspond to the size of the pixels ($\Delta P=0.25^2~\mathrm{arcsec}^2$). Doppler shifts arising from the \NH{} velocity projected along both the star–particle direction and the observer’s line-of-sight are taken into account. The resulting emission from the astrosphere is shown in Figure~\ref{fig: emission_without_ISM_absorption}, where the ISM absorption is not yet included. The first row presents the combined emission from all four \NH{ populations, and the other rows show their separate contributions.}

We present two types of maps: a peak-intensity map (Left column in Figure~\ref{fig: emission_without_ISM_absorption}), which shows the maximum emission in the spectrum, and a total-intensity map (Right column in Figure~\ref{fig: emission_without_ISM_absorption}, which shows the emission integrated over the \lyalpha{} line profile. The brightest structure in the combined peak-intensity map (Figure~\ref{fig: emission_without_ISM_absorption}(a)) is the hydrogen wall, where the \NH{} density is enhanced. Strong emission is also present close to the star, where the incident \lyalpha{} flux from the star is strong. In the combined total-intensity map (Figure~\ref{fig: emission_without_ISM_absorption}(b)), the contribution from near-star regions becomes more prominent.

The combined astrospheric emission is dominated by the \NH{ population generated downstream of the bow shock (Type 1). The original ISM \NH{} (Type 4) also produces strong \lyalpha{} emission, but only in the region outside the astropause. The \NH{} generated downstream of the termination shock (Type 2) contributes primarily to the emission within the astrosphere, whereas the \NH{} in the supersonic stellar wind (Type 3) contributes negligibly.}

\begin{figure}[!htbp]
    \centering
    \includegraphics[width=0.8\linewidth]{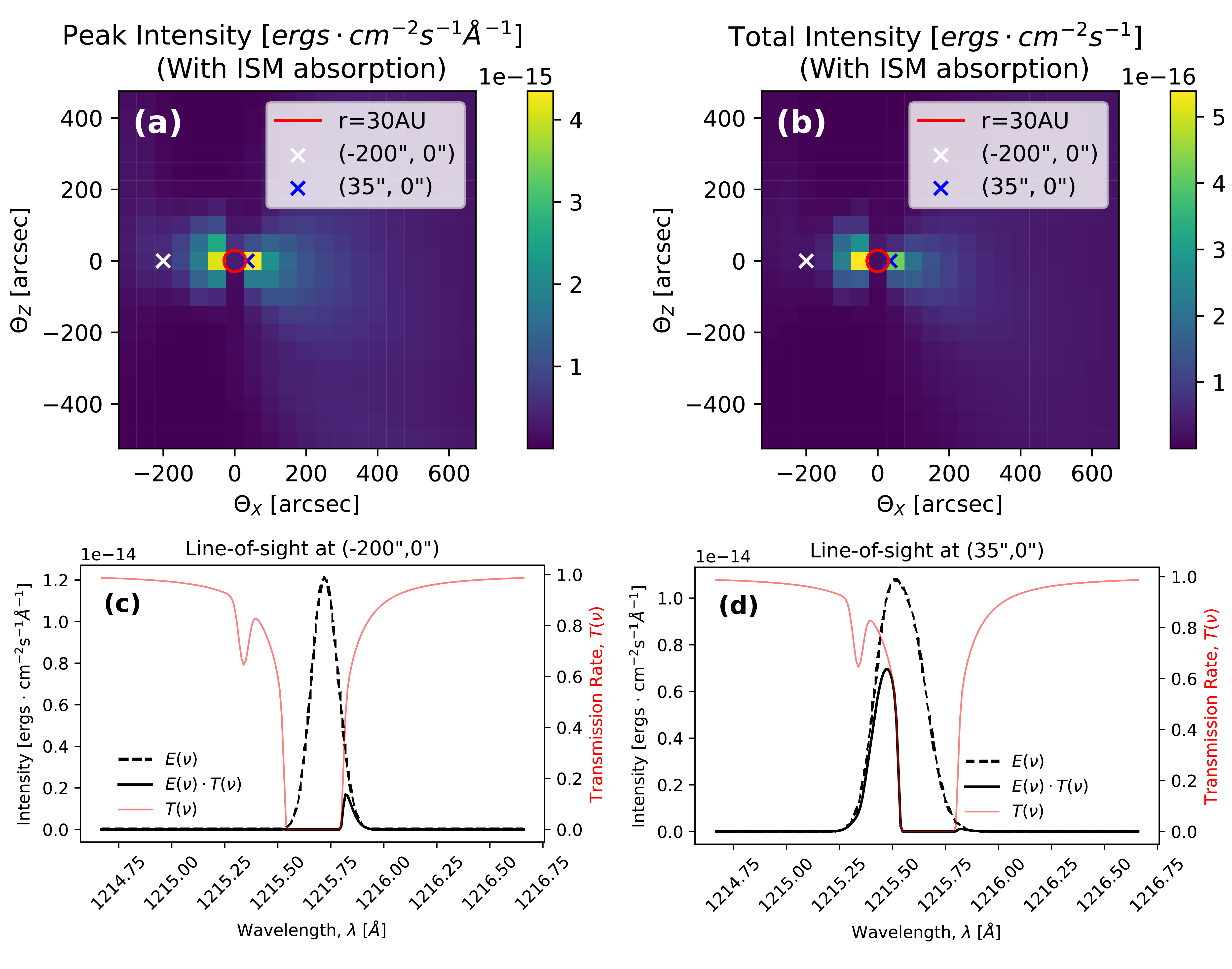}
    \caption{Modeled \lyalpha{} emission from the astrosphere after ISM absorption.
    (a) Distribution of the peak \lyalpha{} intensity.
    (b) \lyalpha{} intensity integrated over wavelength.
    The red circle indicates the inner boundary of the heliosphere model at 30~AU. The white cross marks the line-of-sight through the hydrogen wall, and the blue cross marks the line-of-sight through the near-star stellar wind.
    (c) Original emission $E(\nu)$, ISM transmission $\mathcal{T}(\nu)$, and resulting throughput profile $E(\nu)\cdot \mathcal{T}(\nu)$ along the hydrogen wall line-of-sight.
    (d) Original emission $E(\nu)$, ISM transmission $\mathcal{T}(\nu)$, and resulting throughput profile $E(\nu)\cdot \mathcal{T}(\nu)$ along the near-star line-of-sight.}
    \label{fig: emission_with_ISM_absorption}
\end{figure}

We then include the ISM absorptions in our results by applying the transmission rate. The results are shown in Figure~\ref{fig: emission_with_ISM_absorption}. With ISM absorption, the prominent emission from the hydrogen wall disappears. To illustrate this, we sample two lines of sight separately (Figure~\ref{fig: emission_with_ISM_absorption}(c,d)): one through the hydrogen wall where $\Theta_{X}=-200~\mathrm{arcsec}$ and one within the downwind near-star region where $\Theta_X=35~\mathrm{arcsec}$. The strong central absorption is produced by interstellar neutral hydrogen, while the weaker absorption on the blue side arises from interstellar deuterium. The results show that the peak of hydrogen wall emission coincides with the ISM absorption area, indicating very low throughput. However, the near-star line-of-sight retains a strong peak that survives ISM absorption. This difference arises because the hydrogen wall has a velocity component along the line-of-sight similar to that of the ISM, resulting in strong overlap and absorption, while the stellar wind moves faster along the line-of-sight, producing a blue Doppler shift that shifts its emission away from the ISM absorption center. This feature can be exploited to selectively study \lyalpha{} emission originating inside the astrosphere. We note that the intrinsic solar \lyalpha{} emission (Figure~\ref{fig: OH_model}(b)) is double peaked due to multiple radiative transfer processes of \lyalpha{} photons in a spherically symmetric expanding shell of \NH{} gas \citep{Ahn2004,Verhamme2006}, whereas the emission profiles integrated along the line of sight (Figure~\ref{fig: emission_with_ISM_absorption}(c,d)) consider only single-scattering radiative transfer. This approximation is reasonable because the \NH{} distribution in the astrosphere is expected to be largely laminar.

\begin{figure}[!htbp]
    \centering
    \includegraphics[width=0.8\linewidth]{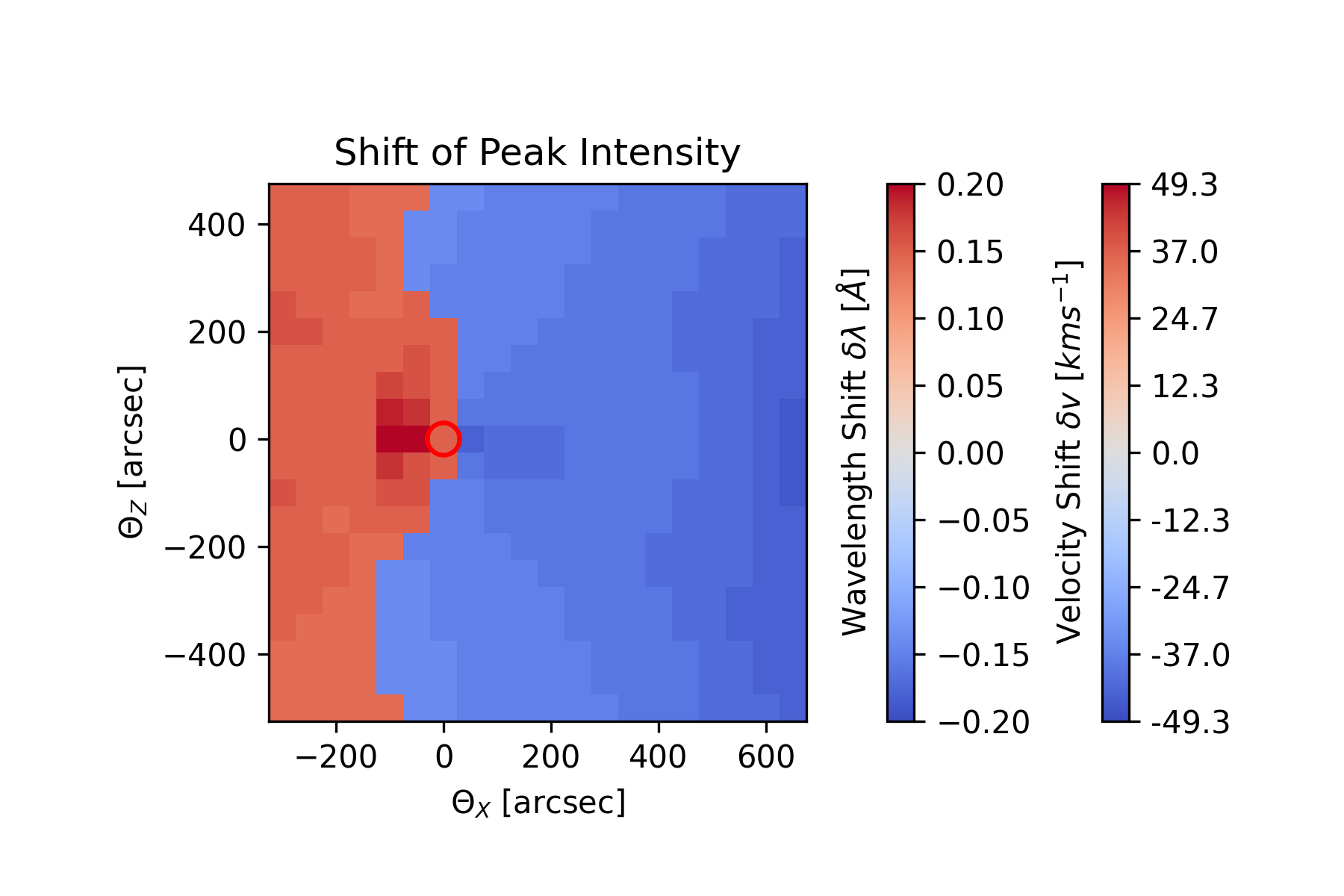}
    \caption{Distribution of the peak positions in the emission spectra after ISM absorption. The positions are given as wavelength shifts $\delta \lambda$ (left colorbar) and the corresponding Doppler velocity shifts $\delta v$ (right colorbar).}
    \label{fig: wavelength_shift}
\end{figure}

We further examine the locations of the emission peaks after ISM absorption. The corresponding wavelength and velocity shifts of the emission peaks are shown in Figure~\ref{fig: wavelength_shift}. In the upwind region—from the star towards the astropause—the emission peaks are redshifted, whereas in the downwind regions inside the astrosphere, the emission peaks are blueshifted. These features arise from Doppler shifts along the star–\NH{} particle lines: photons are emitted isotropically from the star, while the \NH{} particles flow from the upwind to the downwind direction. The peak wavelength changes abruptly rather than gradually because the central part of the emission is absorbed by the ISM. Consequently, astrospheric emission from the downwind side is not expected to be strongly affected by heliospheric absorption, which primarily attenuates the red wing of the line profile.

For a similar observer–astrosphere configuration as assumed here, the locations of the emission peaks can be used to infer several properties of the astrosphere, including the flow direction and the characteristic velocity scale of \NH{} throughout the astrosphere, and can also provide an indicator of the direction of the astropause.

So far, we have calculated the \lyalpha{ emission for a flank view, assuming that the ISM properties are uniform along the line of sight, as in the simulation domain. In reality, however, the relative orientation of the astrosphere may vary, and the ISM properties can change significantly along the line-of-sight. As a result, the relative velocity between the star and the ISM projected along the line of sight may range from $\sim10$ to $\sim100$~km~s$^{-1}$, as inferred from interstellar Mg II observations \citep{WoodandLinsky1998}.}

We further explore 2D emission maps for lines-of-sight with varying angles $\phi$ between the line of sight and the upwind direction of the astrosphere, as well as different relative proper motions along the lines-of-sight between the averaged ISM flow and the star $\Delta V=V_{\mathrm{ISM,~line-of-sight}}-V_{\mathrm{star}}$. We also adopt a Doppler width of $11$~km~s$^{-1}$ for hydrogen \citep{WoodandLinsky1998}. Specifically, we consider three representative orientations: $\phi=45^{\circ}$, where the line-of-sight intersects the hydrogen wall; $\phi=90^{\circ}$, where the line-of-sight is perpendicular to both the ISM flow and the stellar rotation axis, consistent with the previous results; and $\phi=135^{\circ}$, where the line of sight probes the astrotail region. We also examine relative proper motions of $\Delta V=-100$, $-30$, $0$, $30$, and $100$~km~s$^{-1}$.

\begin{figure}
    \centering
    \includegraphics[width=1.0\linewidth]{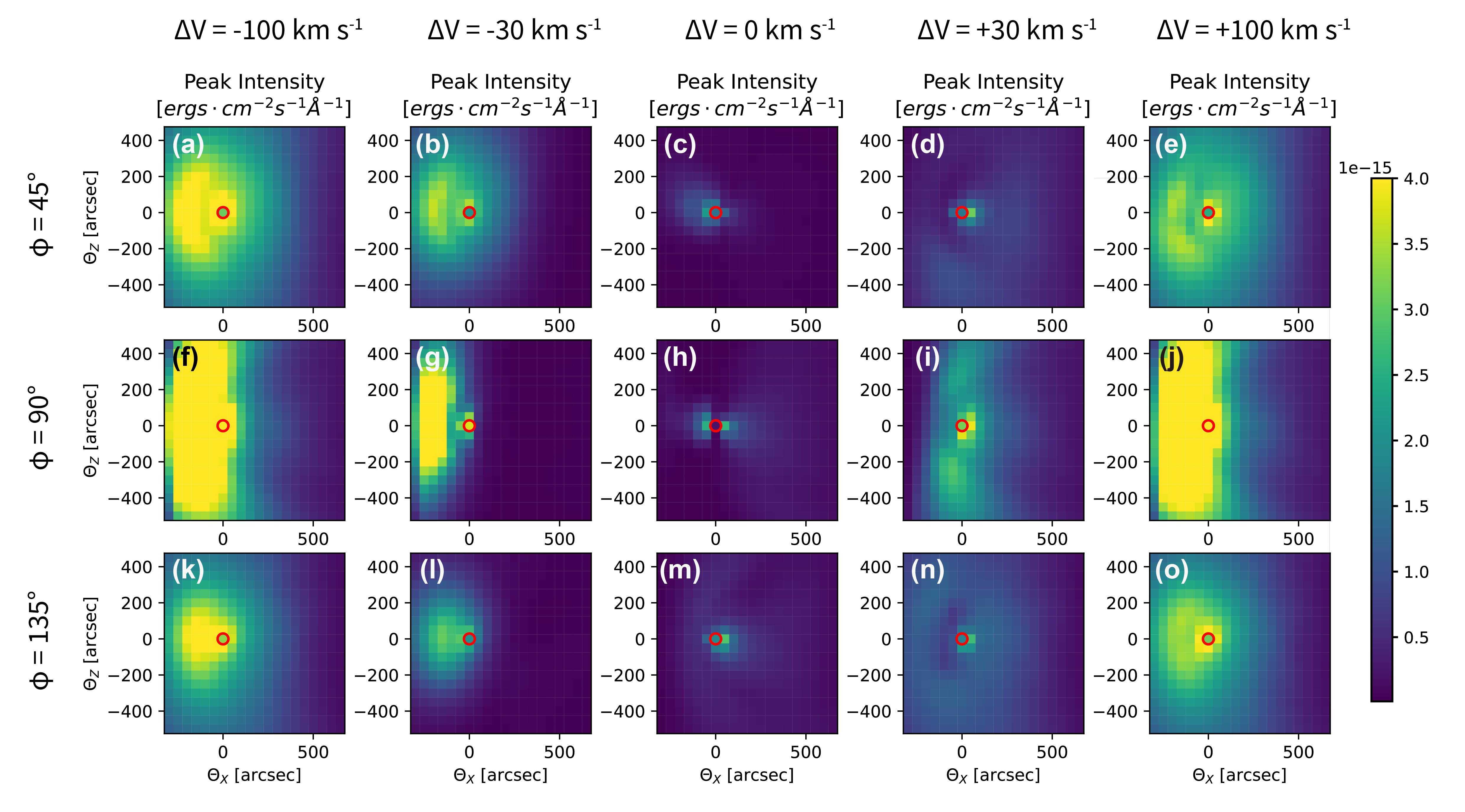}
    \caption{Peak intensity maps for lines-of-sight with different orientations and averaged ISM velocities. From top to bottom, the rows show results with $\phi=45, ~ 90, ~135^{\circ}$, where $\phi$ is the angle between the upwind direction and the line-of-sight. From left to right, the columns show results with $\Delta V=-100,~-30,~0,~30,~100$~km~s$^{-1}$}
    \label{fig: vary_los}
\end{figure}

The results are shown in Figure~\ref{fig: vary_los}. For upwind and tailward viewing geometries, the hydrogen wall appears more curved and located closer to the host star. In the flank view, it becomes more vertical and prominent due to line-of-sight integration through the densest regions of the hydrogen wall. The impact of relative proper motions is also significant. For large relative proper motions ($\Delta V=\pm100$~km~s$^{-1}$), the ISM absorption is well separated from the emission profile, allowing high throughput for both the hydrogen wall and astrospheric emission. We note that panels (f) and (j) are not identical, although they appear similar because of the chosen colorbar range. For smaller relative proper motions ($\Delta V=\pm 30$~km~s$^{-1}$), the effect depends on the sign of $\Delta V$. When $\Delta V>0$, meaning that the ISM flow along the line of sight is directed toward the star, the hydrogen wall emission is strongly absorbed because the neutral hydrogen in the hydrogen wall is also moving towards the star. In contrast, when $\Delta V<0$, the astrospheric emission is preferentially absorbed for the same reason. Additionally, the results show that stars with larger relative proper motions hold stronger emissions for us to observe. 

\section{Detection Prospects}
In our forward modelling, the peak \lyalpha{} intensity can reach $\sim 10^{-15}$ ergs~cm$^{-2}$~s$^{-1}$~$\AA^{-1}$, which is within reach of the detection limit of STIS onboard {\it HST} \citep{STIShandbook2024}. Using the range of line fluxes from our simulations (e.g., Figure~\ref{fig: emission_with_ISM_absorption}), and the STIS exposure time calculator, we evaluate the amount of time needed to reach a signal-to-noise ratio $S/N=5$ with STIS FUV-MAMA in its G140M/c1222 grating configuration. For a flux level of $5\times 10^{-15}$ ergs~cm$^{-2}$~s$^{-1}$~$\AA^{-1}$, the required integration time is approximately 59 minutes, while for a lower flux level of $1\times 10^{-15}$ ergs~cm$^{-2}$~s$^{-1}$~$\AA^{-1}$, the integration time increases to about 8.1 hours. Obviously, multiple STIS slit positions would be needed. To increase the chance of detection, one indeed needs to cover the 2D distribution of astrophere emission (e.g., Figure~\ref{fig: emission_with_ISM_absorption}) as there is no a priori way to estimate the geometry of the emission. While this represents a medium- or large-size program for {\it HST}, detecting and characterising astrosphere emission (flux and geometry) could be one of the prime science goals for the integral-field unit (IFU) far-ultraviolet spectrograph on board the {\it Habitable World Observatory} \citep{Feinberg2026}. Indeed, a 2'$\times$2' IFU size combined with a collective area about an order of magnitude larger than {\it HST} would be able to wage the \lyalpha{} astrosphere emission of nearby solar-type stars in just a few hours of integration.

As a first attempt, our model assumes a heliosphere-like astrosphere located at a distance of 1 pc from the observer. In reality, the nearby stars analyzed by \citet{Wood2021} span distances ranging from approximately 1 pc to several tens of parsecs. Although the total flux from more distant systems decreases as $D^{-2}$, for a pixel of fixed angular size the projected physical area on the plane of the sky increases as $D^2$. Consequently, the flux dilution is compensated, and the surface brightness of the extended astrospheric emission is expected to remain constant. 

Regarding potential targets, one can refer to the stars with confirmed astrospheric absorption in the stellar \lyalpha{} line profile and mass-loss rate estimation, as listed in \citet{Wood2021}. The size of the astrosphere could affect the possibility. Stars with higher mass-loss rate may host larger and denser astrospheres (e.g., $\epsilon$~Eri with distance $D=3.22$~pc, mass-loss rate $\dot{M}=30~\dot{M_{\odot}}$ \citep{Dring1997}), therefore producing stronger emission from the near-star region, but it may be harder to observe astropsheric boundaries. Stars with lower mass-loss rate and higher ISM flow velocity are expected to have a compact astrosphere (e.g. 61~Cyg~A with distance $D=3.48$~pc, mass-loss rate $\dot{M}=0.5 ~\dot{M_{\odot}}$, relative velocity of ISM flow $V_{\mathrm{ISM}}=86$~km~s$^{-1}$ \citep{WoodandLinsky1998}). As a result, the circumstellar \lyalpha{} emission may experience less scattering, potentially making it easier to detect despite the lower mass-loss rate.

\citet{Wood2003} reported a tentative {\it HST}/STIS observation of circumstellar \lyalpha{} emission around 40~Eri~A located 5.0 pc away, but no recognizable signatures were identified. This target is selected due to a combination of a small mass loss rate and strong ISM inflow and is expected to host a compact astrosphere. Using a 2D axisymmetric HD model coupled with radiative transfer calculations, they predicted that the \lyalpha{} emission should have been detectable. The non-detection was attributed to the possibility that the star is embedded in a highly ionized ISM.

Our study adopts a different 3D MHD model combined with a simpler radiative transfer calculation to model the \lyalpha{} emission. Whereas \citet{Wood2003} computed the emission using only the \NH{} distribution in the hydrogen wall, we include all four components of \NH{}. We also explore the influence of different line-of-sight configurations and relative proper motions $\Delta V=V_{\mathrm{ISM}}$ between the ISM and the star. Results suggest that ISM with larger relative proper motions can leave more detectable emissions, while ISM with smaller relative proper motions preferentially absorb hydrogen wall or astrospherical emissions, depending on the direction. Further observations and a broader set of MHD models spanning a range of stellar parameters are required to obtain a clearer picture.

Once equipped with actual observations and a broader set of parametric simulations, it will be possible to retrieve quantitative information and place constraints on astrospheres. For example, the location and shape of the astropause can be inferred, thereby constraining the stellar mass outflow through pressure balance at the astrospheric boundary. In addition, the global symmetry of the astrosphere can be assessed. When combined with stellar magnetogram observations, such analyses may further reveal how the stellar magnetic field influences the overall morphology of the astrosphere.

Finally, the \lyalpha{} emission is subject to both heliospheric absorption and geocoronal contamination. Heliospheric absorption primarily affects the red wing of the interstellar \lyalpha{} absorption profile, because the neutral hydrogen in the heliospheric boundary region is moving away from an observer located near the Sun. Consequently, as long as the peak of the astrospheric \lyalpha{} emission lies on the blue wing of the interstellar absorption, it will not be strongly attenuated by heliospheric absorption.

Geocoronal emission, on the other hand, can usually be well characterized by a narrow Gaussian profile that lies within the interstellar \lyalpha{} absorption band, making it relatively straightforward to identify and remove from observational data \citep{Wood2005}. As demonstrated by \citet{Wood2003}, astrospheric emission that spectrally overlaps with the geocoronal component may be indistinguishable from it. However, such overlap can potentially be minimized or avoided through careful observational planning, specifically by targeting epochs when the relative velocity between the star and the Earth is maximized. We further propose that astrospheres with higher stellar mass-loss rates—and consequently higher stellar wind velocities—may produce \lyalpha{} emission extending beyond the geocoronal wavelength range. Further modelling will be required to quantitatively assess this possibility.

\section{Conclusion}
In this study, we propose a new approach to probe the astrosphere by detecting 2D maps of \lyalpha{} emission surrounding stars. This emission arises from resonant scattering by neutral atoms produced through charge exchange between the stellar wind and the ISM.

We use a 3D MHD heliospheric-like astrosphere model to forward model the resulting \lyalpha{} emission maps. Depending on the relative proper motion $\Delta V$ between the ISM along the line-of-sight and the star, the hydrogen wall and astrospheric emissions might both survive or be partially absorbed. Anyway, the Doppler-shifted features and spatial distributions observed in the 2D \lyalpha{} maps can provide valuable constraints on the stellar wind kinematics and enable a coarse reconstruction of the global astrospheric structure. In particular, such observations can constrain the standoff distance of the bow shock, thereby probing the pressure balance between the stellar wind and the ISM flow, and can further reveal the symmetry and topology of the astro-tail, offering insight into the underlying stellar magnetic field.

We further discuss the observational feasibility of this technique with {\it HST}/STIS and propose some potential targets. This observational strategy could also inform future programs with next-generation facilities such as the {\it Habitable Worlds Observatory} \citep{AA2020}. More modeling work is underway to better quantify the expected outcomes of targeted observations. This approach opens a new avenue for directly imaging astrospheric structures and constraining stellar wind–ISM interactions, and provides a foundation for future observational and theoretical efforts aimed at characterizing stellar winds beyond the solar system.

\section{Acknowledgments}
The authors from China are supported by NSFC, NKRDC, and CNSA under the following grant numbers: 42530105, 42241118, 2021YFA0718600, 2022YFF0503800, 42150105, 42204166, D010301, D050106, and D050103.
Z. Wu is also supported by CSC scholarship No. CSC202406010248. 
TVD was supported by a Senior Research Project (G088021N) of the FWO Vlaanderen. Furthermore, H.S, TVD and Z. Wu received financial support from the Flemish Government under the long-term structural Methusalem funding program, project SOUL: Stellar evolution in full glory, grant METH/24/012 at KU Leuven. NJ acknowledge founding from the BELgian federal Science Policy Office (BELSPO) through PRODEX grant PLATO: ZKE2050-01-D01.

The research that led to these results was subsidised by the Belgian Federal Science Policy Office through the contract B2/223/P1/CLOSE-UP and through the PRODEX grant Plato: ZKE2050-01-D01. It is also part of the DynaSun project and has thus received funding under the Horizon Europe programme of the European Union under grant agreement (no. 101131534). Views and opinions expressed are however those of the author(s) only and do not necessarily reflect those of the European Union and therefore the European Union cannot be held responsible for them. 

The authors would like to thank the referee, Dr.~Linsky, for his valuable comments, which greatly improved this paper.


\end{document}